\documentclass[singlecolumn]{revtex4}
\usepackage{graphicx}

\def\textbf#1{{\bf #1}}

\def\be{\begin{equation}}
\def\ee{\end{equation}}

\def\beqa{\begin{eqnarray}}
\def\eeqa{\end{eqnarray}}
\begin{document}

\title{Nonclassicality of quantum excitation of classical coherent fields in thermal environments}

\author{Shang-Bin Li$^{1}$}, \author{Justin Liu$^{1}$}, \author{Xu-Bo Zou$^{2}$}, \author{Guang-Can Guo$^{2}$}

\affiliation{1. Shanghai research center of Amertron-global,
Zhangjiang High-Tech Park, \\
299 Lane, Bisheng Road, No. 3, Suite 202, Shanghai, 201204, P.R.
China.} \affiliation{2. Key Laboratory of Quantum Information,
University of Science and Technology of China, Hefei 230026,
China.}
\begin{abstract}
{\normalsize The nonclassicality of photon-added coherent fields in
the thermal channel is investigated by exploring the volume of the
negative part of the Wigner function which reduces with the
dissipative time. The Wigner functions become positive when the
decay time exceeds a threshold value. For the case of the single
photon-added coherent state, we derive the exact threshold values of
decay time in the thermal channel. For arbitrary partial negative
Wigner distribution function, a generic analytical relation between
the mean photon number of heat bath and the threshold value of decay
time is presented. Finally, the possible application of SPACSs in
quantum computation has been briefly discussed.

OCIS codes: 270.0270, 270.2500, 000.5490}

\end{abstract}

\maketitle

\section{INTRODUCTION}

Nonclassical optical fields play a crucial role in understanding
fundamentals of quantum physics and have many applications in
quantum information processing \cite{Bouwmeester}. Recently, the
preparations of nonclassical non-gaussian optical fields have
attracted much attention. Usually, the nonclassicality manifests
itself in specific properties of quantum statistics, such as the
antibunching \cite{Kimble}, sub-poissonian photon statistics
\cite{Short}, squeezing in one of the quadratures of the field
\cite{Dodonov}, partial negative Wigner distribution
\cite{Hillery}, etc.. The interaction between the nonclassical
optical fields and their surrounding environment causes the
dissipation or dephasing \cite{Gardiner}, which ordinarily
deteriorates the degree of nonclassicality. The Wigner function is
a quasi-probability distribution, which fully describes the state
of a quantum system in phase space. The partial negativity of the
Wigner function is indeed a good indication of the highly
nonclassical character of the optical fields. Reconstruction of
the Wigner distribution in experiments with quantum tomography
\cite{Vogel,Smithey,Welsch} demonstrates appearance of its
negative values. This peculiarity can not be explained in the
framework of the probability theory, hence it does not have any
classical counterparts. Thus, to seek certain relations between
the partial negativity of the Wigner distribution and the
intensity of thermal noise may be very desirable for
experimentally quantifying the variation of nonclassicality of
optical fields in thermal channels. The volume of negative part of
the Wigner function has been proposed as a useful tool for
characterizing the nonclassical optical fields
\cite{Benedict1999,Dodonov2003,Kenfack2004}.

The photon-added coherent states (PACSs) were introduced by
Agarwal and Tara \cite{Agarwal1991}. The single photon-added
coherent state (SPACS) has been experimentally prepared by Zavatta
et al. and its nonclassical properties have been detected by
homodyne tomography technology \cite{Zavatta2004,Zavatta2005}.
Such a state represents the intermediate non-Gaussian state
between quantum Fock state and classical coherent state (with
well-defined amplitude and phase) \cite{Glauber}. For the SPACS, a
quantum to classical transition has been explicitly demonstrated
by ultrafast time-domain quantum homodyne tomography technique.
Previous numerical study has indicated that the partial negativity
of the Wigner function of photon-added coherent states in photon
loss channel will completely disappear as the decay time exceeds a
threshold value \cite{Li2007}. However, the exact threshold value
of the decay time has not been explicitly given. In this paper, we
further investigate the nonclassicality of photon-added coherent
states in the thermal channel by exploring the partial negative
Wigner distribution. The volume of the negative part of the Wigner
function is analyzed. The threshold values of the decay time
corresponding to the transition of the Wigner distribution from
partial negative to completely positive are derived. For SPACS in
thermal channel, it is shown that the threshold value of the decay
time is independent of the initial seed beam intensity. In
addition, for any initial partial negative Wigner functions in
thermal channel, we also obtain a general formula about the
threshold value of decay time which is given by
$\gamma{t}_c(n)=\ln\frac{e^{\gamma{t}_c(0)}+2n}{1+2n}$, where
$\gamma{t}_c(n)$ denotes the threshold value of the decay time for
the case of thermal channel with mean thermal photon number $n$.
From this generic expression, we can clarify how the thermal noise
shortens the threshold value of the decay time compared with the
case of photon-loss.

This paper is organized as follows: In Sec.II, we briefly outline
the basic contents of the photon-added coherent states. The explicit
analytical expression of time evolution Wigner function of the SPACS
in the thermal channel is derived. In Sec.III, the dynamical
behaviors of the volume of the negative part of the Wigner
distribution for the SPACS and the two-photon added coherent states
(TPACSs) is numerically calculated. It is shown that the volume of
the negative part decreases with the decay time. The volume of the
negative part of Wigner function of the TPACS more rapidly decreases
than the one of the SPACS, though their Wigner functions become
non-negative at the same threshold decay time. A generic relation
between the threshold decay time of arbitrary partial negative
Wigner functions in the thermal channel and the mean thermal photon
number of the thermal reservoir is presented. Moreover, the possible
application of SPACSs in quantum computation has been briefly
discussed. In Sec.IV, there are some concluding remarks.

\section{WIGNER FUNCTIONS OF PHOTO-ADDED COHERENT STATES IN THERMAL ENVIRONMENT}

Let us first briefly recall the definition of the photon-added
coherent states (PACSs) \cite{Agarwal1991}. The PACSs are defined
by $\frac{1}{\sqrt{N(\alpha,m)}}a^{\dagger{m}}|\alpha\rangle$,
where $|\alpha\rangle$ is the coherent state with the amplitude
$\alpha$ and $a^{\dagger}$ ($a$) is the creation operator
(annihilation operator) of the optical mode.
$N(\alpha,m)=m!L_m(-|\alpha|^2)$, where $L_m(x)$ is the
$m$th-order Laguerre polynomial. When the PACS evolves in the
thermal channel, the evolution of the density matrix can be
described by \cite{Gardiner} \beqa
\frac{d\rho}{dt}&=&\frac{\gamma(n+1)}{2}(2a\rho{a}^{\dagger}-a^{\dagger}a\rho-\rho{a}^{\dagger}a)\nonumber\\
&&+\frac{\gamma{n}}{2}(2a^{\dagger}\rho{a}-aa^{\dagger}\rho-\rho{a}{a}^{\dagger}),
\eeqa where $\gamma$ represents dissipative coefficient and $n$
denotes the mean thermal photon number of the heat bath. When
$n=0$, the Eq.(1) reduces to the master equation describing the
photon-loss channel.

The presence of negativity in the Wigner function of the optical
field is the indicator of nonclassicality. For an optical field in
the state $\rho$, the Wigner function, the Fourier transformation
of characteristics function \cite{Barnett} of the state can be
derived by \cite{Cessa,Englert} \be
W(\beta)=\frac{2}{\pi}{\mathrm{Tr}}[(\hat{O}_e-\hat{O}_o)\hat{D}(\beta)\rho\hat{D}^{\dagger}(\beta)],\ee
where $\hat{O}_e\equiv\sum^{\infty}_{n=0}|2n\rangle\langle2n|$ and
$\hat{O}_o\equiv\sum^{\infty}_{n=0}|2n+1\rangle\langle2n+1|$ are
the even and odd parity operators respectively. In the thermal
channel described by the master equation (1), the time evolution
Wigner function satisfies the following Fokker-Planck equation
\cite{Carmichael} \beqa
\frac{\partial}{\partial{t}}W(q,p,t)&=&\frac{\gamma}{2}(\frac{\partial}{\partial{q}}q+\frac{\partial}{\partial{p}}p)W(q,p,t)\nonumber\\
&&+\frac{\gamma(2n+1)}{8}(\frac{\partial^2}{\partial{q}^2}+\frac{\partial^2}{\partial{p}^2})W(q,p,t).\eeqa
where $q$ and $p$ represent the real part and imaginary part of
$\beta$, respectively. The time evolution Wigner function can be
derived as following: \be
W(q,p,\gamma{t})=\exp(\gamma{t})\int^{\infty}_{-\infty}\int^{\infty}_{-\infty}W_T(x,y)W(\frac{q-\sqrt{1-e^{-\gamma{t}}}x}{\sqrt{e^{-\gamma{t}}}},\frac{p-\sqrt{1-e^{-\gamma{t}}}y}{\sqrt{e^{-\gamma{t}}}},0)dxdy\ee
where \be
W_T(x,y)=\frac{2}{\pi(1+2n)}\exp(-\frac{2(x^2+y^2)}{1+2n})\ee is
the Wigner function of the thermal state with mean photon number
$n$. Substituting the initial Wigner function of the pure SPACS
\cite{Agarwal1991} \be
W^S(q,p,0)=\frac{-2L_1(|2q+2ip-\alpha|^2)}{\pi{L_1}(-|\alpha|^2)}e^{-2|q+ip-\alpha|^2}\ee
and the Wigner function of the pure TPACS \cite{Agarwal1991} \be
W^T(q,p,0)=\frac{2L_2(|2q+2ip-\alpha|^2)}{\pi{L_2}(-|\alpha|^2)}e^{-2|q+ip-\alpha|^2}\ee
into the Eq.(4), we can obtain the corresponding time evolution
Wigner functions. For the case of SPACS, the analytical time
evolution Wigner function can be obtained \beqa
W^S(q,p,\gamma{t})&=&\frac{2e^{\gamma{t}}[(\xi-c^2{\mathrm{Re}}\alpha)^2+(\zeta-c^2{\mathrm{Im}}\alpha)^2+c^4-1]\exp[-2(\mu^2+\nu^2)/(1+c^2)]}{\pi(1+|\alpha|^2)(1+c^2)^3},\nonumber\\
c&=&[(\exp(\gamma{t})-1)(1+2n)]^{1/2},\nonumber\\
\mu&=&{\mathrm{Re}}(\alpha)-q\exp(\gamma{t}/2),\nonumber\\
\nu&=&{\mathrm{Im}}(\alpha)-p\exp(\gamma{t}/2),\nonumber\\
\xi&=&{\mathrm{Re}}(\alpha)-2q\exp(\gamma{t}/2),\nonumber\\
\zeta&=&{\mathrm{Im}}(\alpha)-2p\exp(\gamma{t}/2). \eeqa  In
Fig.1, the Wigner function of the SPACS with $\alpha=0.5$ in the
thermal channel with $n=1$ at two different values of decay time
are plotted. It is shown that the phase space Wigner distribution
of the pure SPACS with $\alpha=0.5$ loses its circular symmetry
and moves away from the origin due to the appearance of a definite
phase. The partial negativity of the Wigner function indicates the
nonclassical nature of the single quantum excitation of the
classical coherent field. The thermal noise causes the
disappearance of the partial negativity of the Wigner function if
the decay time exceeds a threshold value. The tilted ringlike
wings in the distribution gradually start to disappear and the
distribution becomes more and more similar to the Gaussian typical
of a thermal state. Eq.(8) also indicates that the negative region
of the Wigner function of the slightly decayed SPACS is always a
circle. Fig.2 illustrates the phase space Wigner distributions of
the TPACS with $\alpha=0.5$ in the thermal channel with $n=1$. The
upper figure of Fig.2 shows the region of the negative part of the
Wigner function of the pure TPACS is a ring belt which is
different from the case of SPACSs. The lower figure of Fig.2 shows
that the absolute value of negative minimum of the Wigner
distribution decreases as $\gamma{t}$ increases. When $\gamma{t}$
exceeds a threshold value, the partial negativity of the Wigner
distribution of the TPACS also completely disappears.

\section{INFLUENCE OF THERMAL NOISE ON THE NONCLASSICALITY OF QUANTUM EXCITATION OF CLASSICAL COHERENT FIELDS}

Recently, the volume $P_{NW}$ of the negative part of the Wigner
function has been suggested as a good choice for quantifying the
nonclassicality \cite{Benedict1999,Dodonov2003,Kenfack2004}.
$P_{NW}$ is defined by \be P_{NW}=|\int_{\Omega}W(q,p)dqdp|,\ee
where $\Omega$ is the negative Wigner distribution region. In
Ref.\cite{Li2007}, we have investigated $P_{NW}$ of SPACS and
TPACS in the photon-loss channel. It was shown that $P_{NW}$ and
entanglement potential defined in Ref.\cite{Asboth2005} exhibit
the consistent behaviors in short decay time.

In this section, we bring our attention to the influence of
thermal noise on the nonclassicality of the quantum excitation of
classical coherent optical fields by calculating $P_{NW}$. In
Eq.(8), we have obtained the analytical solution of the time
evolving Wigner function of the SPACS in the thermal channel.
Following similar lines, we can derive the one of the TPACS too.
But the expressions are lengthy and do not exhibit a simple
structure.

Based on the analytical expressions of the Wigner functions, we
have calculated the volume of the negativity in Wigner functions
for both the SPACSs and the TPACSs. In Fig.3, $P_{NW}$ of SPACSs
and TPACSs with the parameter $\alpha=1.5$ is plotted as the
function of $\gamma{t}$ for nine different values of the mean
thermal photon number $n$ of the thermal channel. It is shown that
the thermal noise deteriorates the partial negativity, and
$P_{NW}$ monotonically decreases with the decay time. The larger
the thermal noise, the more rapidly $P_{NW}$ decreases, which
implies that the nonclassicality of the optical fields are very
fragile against the thermal noise. From Fig.4, it can be observed
that, the volume of the negative part of the TPACS's Wigner
distribution is more fragile than the one of the SPACS's Wigner
distribution against the thermal noise if the seed beam intensity
$|\alpha|^2$ is not very large, though the initial pure TPACSs
have larger values of $P_{NW}$ than the pure SPACSs with the same
value of $\alpha$.

The above results also indicate that $P_{NW}$ becomes zero at a
threshold decay time $\gamma{t}_c$ which depends on the value of
$n$. For the case of the SPACS in thermal channel, we can directly
derive the threshold decay time from the Eq.(8). $\gamma{t}_c$ can
be obtained as follows: \be \gamma{t}_c=\ln(\frac{2+2n}{1+2n}),\ee
which shows the threshold decay time is independent of the seed beam
intensity $|\alpha|^2$ of the SPACSs. For the case of the TPACSs,
the threshold decay times also satisfy the Eq.(10). In Fig.5, we
have plotted both the exact analytical and the numerical results of
the threshold decay time as the function of $n$ for the case of
SPACSs with $\alpha=0.5$. Well consistent between the analytical
results and the numerical solutions is found.

For arbitrary nonclassical optical fields which have the partial
negative Wigner distribution function, there exists a relation
between the mean photon number $n$ of the thermal reservoir and
the threshold decay time $\gamma{t}_c(n)$ beyond which their
Wigner function become positive. \be
\gamma{t}_c(n)=\ln\frac{e^{\gamma{t}_c(0)}+2n}{1+2n},\ee where
$\gamma{t}_c(0)$ is the threshold decay time in the photon loss
channel. In the derivation of the Eq.(11), we have reformulated
Eq.(4) as \be
W(q,p,\gamma{t})=\exp(\gamma{t})\int^{\infty}_{-\infty}\int^{\infty}_{-\infty}W_0(x^{\prime},y^{\prime})W(e^{\gamma{t^{\prime}}/2}q^{\prime}-\sqrt{e^{\gamma{t^{\prime}}}-1}x^{\prime},e^{\gamma{t^{\prime}}/2}p^{\prime}-\sqrt{e^{\gamma{t^{\prime}}}-1}y^{\prime},0)dx^{\prime}dy^{\prime},\ee
where \beqa
W_0(x^{\prime},y^{\prime})&=&\frac{2}{\pi}\exp(-2(x^{\prime2}+y^{\prime2}))\nonumber\\
x^{\prime}&=&\frac{x}{\sqrt{1+2n}}\nonumber\\
y^{\prime}&=&\frac{y}{\sqrt{1+2n}}\nonumber\\
q^{\prime}&=&\frac{e^{\gamma{t}/2}}{\sqrt{1+(1+2n)(e^{\gamma{t}}-1)}}q\nonumber\\
p^{\prime}&=&\frac{e^{\gamma{t}/2}}{\sqrt{1+(1+2n)(e^{\gamma{t}}-1)}}p\nonumber\\
\gamma{t}^{\prime}&=&\ln[1+(1+2n)(e^{\gamma{t}}-1)]. \eeqa
Obviously, the right side of the Eq.(12) represents a scaled time
evolution Wigner function in the photon loss channel, i.e. $n=0$.
Therefore we have \be
W(q,p,\gamma{t})/e^{\gamma{t}}=W^{(0)}(q^{\prime},p^{\prime},\gamma{t^{\prime}})/e^{\gamma{t^{\prime}}},\ee
where $W^{(0)}(q,p,\gamma{t})$ is the time evolution Wigner function
of the optical field in the photon loss channel. Based on Eq.(14)
and the definition of $P_{NW}$ in Eq.(9), we can prove \be
P_{NW}(\gamma{t})=P^{(0)}_{NW}(\gamma{t}^{\prime}), \ee where
$P^{(0)}_{NW}(\gamma{t}^{\prime})$ denotes the volume of the
negative part of the photon-loss-induced time evolving Wigner
function at the time ${t}^{\prime}$ of the optical field initially
described by the wigner function $W(q,p,0)$. Assuming we have known
the threshold decay time of arbitrary partial negative Wigner
function in the photon loss channel, we can derive the corresponding
threshold decay time of that Wigner function in the thermal channel
via the Eq.(15). If $P^{(0)}_{NW}(\gamma{t}^{\prime})$ becomes zero
at the decay time $\gamma{t}^{\prime}=\gamma{t}_c(0)$, it can be
found that $P_{NW}(\gamma{t})$ becomes zero at
$\gamma{t}=\ln\frac{e^{\gamma{t}_c(0)}+2n}{1+2n}$ according to the
relation $\gamma{t}^{\prime}=\ln[1+(1+2n)(e^{\gamma{t}}-1)]$. Thus,
we finish proving the relation in Eq.(11) for arbitrary partial
negative Wigner function. This relation may have some potential
applications in the measurement of the temperature of the
surrounding heat bath. In Ref.\cite{Li2007}, we have numerically
investigated the entanglement potential and partial negativity of
the Wigner function of SPACSs in photon-loss channel. The
nonclassical properties of single-photon subtracted squeezed vacuum
states in amplitude decay channel or in the phase diffusion channel
have also been studied by Biswas and Agarwal \cite{Biswas}. The
photon loss channel is only a specific example of the thermal
channel at which the thermal reservoir is at zero-temperature.
Present results not only generalize the results in
Ref.\cite{Li2007}, and give out an analytical expression of the
evolving Wigner function of SPACSs in thermal channel, but also find
out the exact analytical relation between the mean thermal photon
number of the thermal channel and the threshold time concerning the
complete disappearance of partial negativity of the Wigner function
of SPACSs. Furthermore, for any initial partial negative Wigner
distribution function in thermal channel, we derive a general
relation between the threshold decay time and the mean thermal
photon number of the thermal channel. Roughly speaking, the formula
in Eq.(11) implies that the higher the temperature of thermal
channel, the more rapidly the nonclassicality of the optical fields
with partial negative Wigner distribution decays. From Eq.(11), it
is also easy to know that
$\gamma{t}_c(n)\simeq\frac{e^{\gamma{t}_c(0)}-1}{2n}$ when $n\gg1$,
and $\gamma{t}_c(n)\simeq\gamma{t}_c(0)-2(1-e^{-\gamma{t}_c(0)})n$
when $n\ll1$.

In the past five years, much attention has been focused on the
application of coherent states in quantum computation
\cite{Ralph2003,Ralph2002}. SPACSs can also find many applications
in quantum information processes such as the quantum computation.
Similar to the schemes in the literatures, one can regard the vacuum
state $|0\rangle$ and the SPACS
$\frac{1}{\sqrt{N(\alpha,1)}}a^{\dagger}|\alpha\rangle$ as two
orthogonal states of an optical qubit $|0\rangle_L$ and
$|1\rangle_L$, respectively. As an illustration, in this state
encoding, a nontrivial two qubit gate can be implemented using only
a beam splitter. Consider the beam splitter interaction described by
the unitary transformation \be
U_{BS}=\exp[i\phi(ab^{\dagger}+a^{\dagger}b)], \ee where $a$ and $b$
represent the annihilation operators corresponding to two qubits
$|\kappa\rangle_a$ and $|\chi\rangle_b$ which take the states
$|0\rangle$ or
$\frac{1}{\sqrt{N(\alpha,1)}}a^{\dagger}|\alpha\rangle$. It is easy
to derive that the output state produced by such an interaction is
\beqa
U_{BS}|0\rangle_a|0\rangle_b&=&|0\rangle_a|0\rangle_b,\nonumber\\
U_{BS}\frac{1}{\sqrt{N(\alpha,1)}}a^{\dagger}|\alpha\rangle_a|0\rangle_b&=&\frac{1}{\sqrt{N(\alpha,1)}}(\cos(\phi)a^{\dagger}+i\sin(\phi)b^{\dagger})|\cos(\phi)\alpha\rangle_a|i\sin(\phi)\alpha\rangle_b,\nonumber\\
U_{BS}\frac{1}{\sqrt{N(\alpha,1)}}b^{\dagger}|0\rangle_a|\alpha\rangle_b&=&\frac{1}{\sqrt{N(\alpha,1)}}(\cos(\phi)b^{\dagger}+i\sin(\phi)a^{\dagger})|i\sin(\phi)\alpha\rangle_a|\cos(\phi)\alpha\rangle_b,\nonumber\\
U_{BS}\frac{1}{N(\alpha,1)}a^{\dagger}b^{\dagger}|\alpha\rangle_a|\alpha\rangle_b&=&\frac{1}{N(\alpha,1)}(\cos(\phi)a^{\dagger}+i\sin(\phi)b^{\dagger})(\cos(\phi)b^{\dagger}+i\sin(\phi)a^{\dagger})|e^{i\phi}\alpha\rangle_a|e^{i\phi}\alpha\rangle_b,
\eeqa The overlap between the output and input states can be
expressed as follows: \beqa
{}_a\langle0|{}_b\langle0|U_{BS}|0\rangle_a|0\rangle_b=1,\nonumber\\
\frac{1}{N(\alpha,1)}{}_a\langle\alpha|{}_b\langle0|aU_{BS}a^{\dagger}|\alpha\rangle_a|0\rangle_b\nonumber\\
=\frac{1}{N(\alpha,1)}{}_a\langle0|{}_b\langle\alpha|bU_{BS}b^{\dagger}|0\rangle_a|\alpha\rangle_b\nonumber\\
=\frac{\cos^2(\phi)|\alpha|^2+\cos\phi}{N(\alpha,1)}e^{|\alpha|^2(\cos\phi-1)},\nonumber\\
\frac{1}{N^2(\alpha,1)}{}_a\langle\alpha|{}_b\langle\alpha|abU_{BS}a^{\dagger}b^{\dagger}|\alpha\rangle_a|\alpha\rangle_b\nonumber\\
=\frac{|\alpha|^4e^{4i\phi}+2|\alpha|^2e^{3i\phi}+\cos(2\phi)}{N^2(\alpha,1)}e^{2|\alpha|^2(e^{i\phi}-1)}
\eeqa If $\phi$ is assumed to be sufficiently small such that
$\phi^2|\alpha|^2\ll1$ but $|\alpha|$ is sufficiently large that
$\phi|\alpha|^2$ is of order one. Then Eq.(18) can approximately
become  \beqa
{}_a\langle0|{}_b\langle0|U_{BS}|0\rangle_a|0\rangle_b=1,\nonumber\\
\frac{1}{N(\alpha,1)}{}_a\langle\alpha|{}_b\langle0|aU_{BS}a^{\dagger}|\alpha\rangle_a|0\rangle_b\nonumber\\
=\frac{1}{N(\alpha,1)}{}_a\langle0|{}_b\langle\alpha|bU_{BS}b^{\dagger}|0\rangle_a|\alpha\rangle_b\nonumber\\
\approx1,\nonumber\\
\frac{1}{N^2(\alpha,1)}{}_a\langle\alpha|{}_b\langle\alpha|abU_{BS}a^{\dagger}b^{\dagger}|\alpha\rangle_a|\alpha\rangle_b\nonumber\\
\approx{e}^{2i\phi|\alpha|^2}. \eeqa Furthermore, the condition
$2\phi|\alpha|^2=\pi$ will guarantee the realization of a controlled
phase gate by the beam splitter acted on the above encoding space.

The results in this paper may be useful for investigating the
influence of the thermal noise on the quantum information processes
in which the photon added coherent state is used as a resource. It
is also interesting to investigate the fidelity of the above
controlled phase gates in thermal channel. The details will be
discussed elsewhere.

\section{CONCLUSIONS}

In summary, we have investigated the non-classicality of photon
excitation of classical coherent field in the thermal channel by
exploring the partial negativity of the Wigner function. The total
volume of the negative part defined by the absolute value of the
integral of the Wigner function over the negative distribution
region is calculated and the partial negativity of the Wigner
function can not be observed when the decay time exceeds a threshold
value which depends on the mean thermal photon number. For the cases
in which the seed beam intensity $|\alpha|^2$ is not very large, it
is found that the TPACSs more rapidly lose their partial negativity
of the Wigner distribution function than the SPACSs with the same
seed beam intensity. For the case of SPACSs in thermal channel, the
exact threshold value of the decay time beyond which the evolving
Wigner function becomes positive is given as
$\gamma{t}_c=\ln(\frac{2+2n}{1+2n})$. For the arbitrary nonclassical
optical fields with partial negative Wigner function, we also
present a generic relation between the threshold decay time of the
thermal channel and the mean thermal photon number of the thermal
reservoir under the assumption of the known threshold decay time of
the photon-loss channel. Finally, the possible application of SPACSs
in quantum computation has been briefly discussed. Present results
in this paper may be useful for checking the influence of the
thermal noise on the quantum information processes in which the
photon added coherent state is used as a resource.




\section * {ACKNOWLEDGMENTS}

This work was supported by National Fundamental Research Program,
also by National Natural Science Foundation of China (Grant No.
10674128 and 60121503) and the Innovation Funds and
\textquotedblleft Hundreds of Talents\textquotedblright\ program
of Chinese Academy of Sciences and Doctor Foundation of Education
Ministry of China (Grant No. 20060358043)

\newpage

\section * {List of Figure Captions}

Fig.1. The Wigner functions of the SPACS with $\alpha=0.5$ in
thermal channel with $n=1$ are depicted for two different values
of decay time $\gamma{t}$.

\noindent Fig.2. The Wigner functions of the TPACS with
$\alpha=0.5$ in thermal channel with $n=1$ are depicted for two
different values of decay time $\gamma{t}$.

\noindent Fig.3. The $P_{NW}$ of the SPACS and TPACS with
$\alpha=1.5$ in thermal channel are depicted as the function of
$\gamma{t}$ for different values of mean thermal photon number
$n$. From top to bottom, $n=0.1,0.2,0.3,0.4,0.5,0.6,0.7,0.8,0.9$.

\noindent Fig.4. $P_{NW}$ is plotted as the function of the decay
time $\gamma{t}$ for SPACS and TPACS with different values of
$\alpha$. (Solid line) from top to bottom, SPACS with
$\alpha=0.1,0.5,1.0,1.5$, respectively; (Dash line) from top to
bottom, TPACS with $\alpha=0.1,0.5,1.0,1.5$, respectively.
$n=0.5$.

\noindent Fig.5. The threshold decay time $\gamma{t}_c$ beyond
which $P_{NW}=0$ is plotted as the function of the mean thermal
photon number $n$ for the case of the SPACS. (Solid square)
numerical results; (Solid line)
$\gamma{t}_c=\ln\frac{2+2n}{1+2n}$. It is shown that
$\gamma{t}_c\simeq0.5/n$ when $n\gg1$. The numerical calculations
are based on the SPACS with $\alpha=0.5$.


\newpage
 \begin{figure}
\centerline{\includegraphics[width=8.3cm]{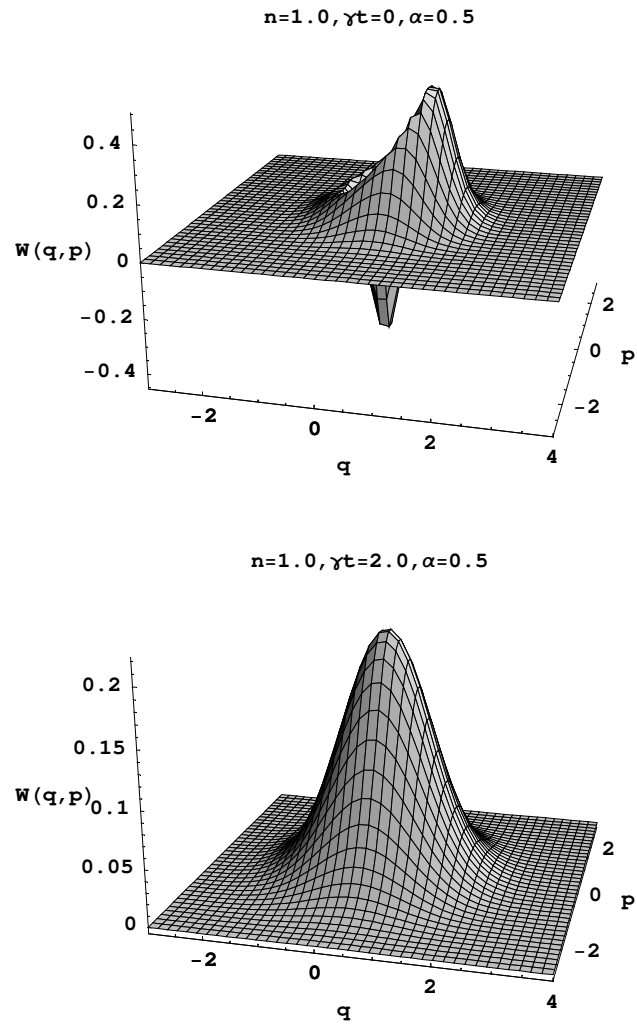}}
\caption{The Wigner functions of the SPACS with $\alpha=0.5$ in
thermal channel with $n=1$ are depicted for two different values
of decay time $\gamma{t}$.}
\end{figure}
\begin{figure}
\centerline{\includegraphics[width=8.3cm]{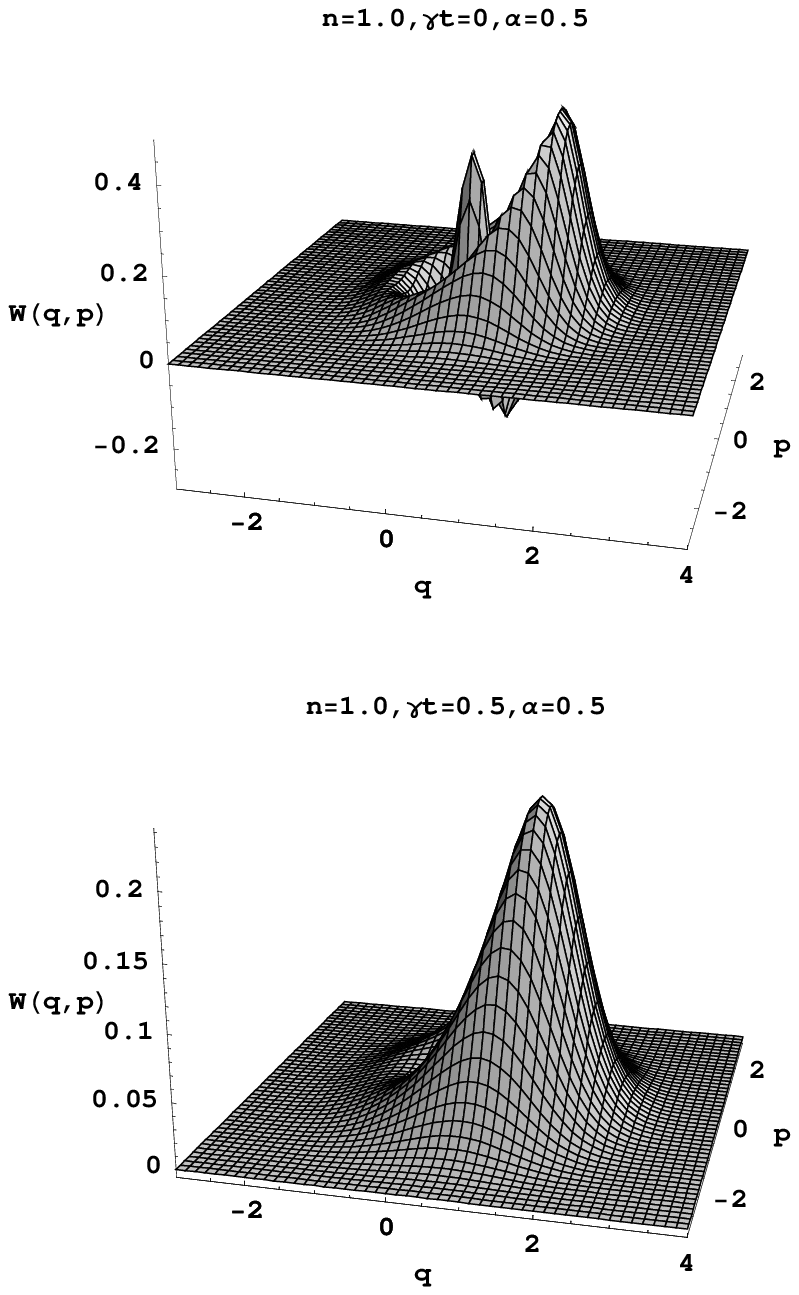}}
\caption{The Wigner functions of the TPACS with $\alpha=0.5$ in
thermal channel with $n=1$ are depicted for two different values
of decay time $\gamma{t}$.}
\end{figure}
\begin{figure}
\centerline{\includegraphics[width=8.3cm]{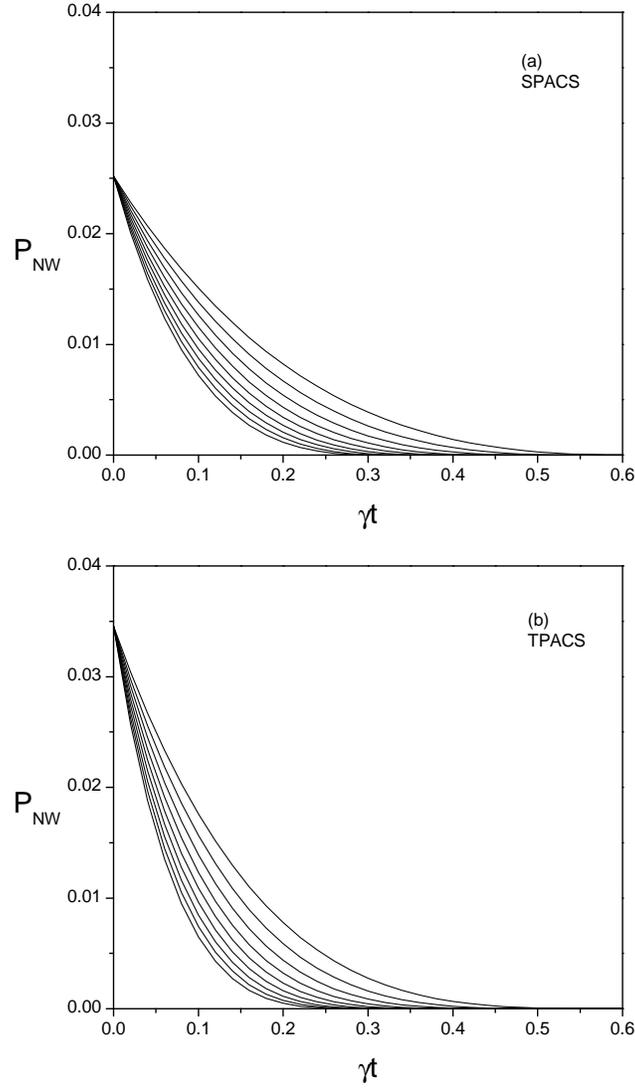}}
\caption{The $P_{NW}$ of the SPACS and TPACS with $\alpha=1.5$ in
thermal channel are depicted as the function of $\gamma{t}$ for
different values of mean thermal photon number $n$. From top to
bottom, $n=0.1,0.2,0.3,0.4,0.5,0.6,0.7,0.8,0.9$.}
\end{figure}
\begin{figure}
\centerline{\includegraphics[width=8.3cm]{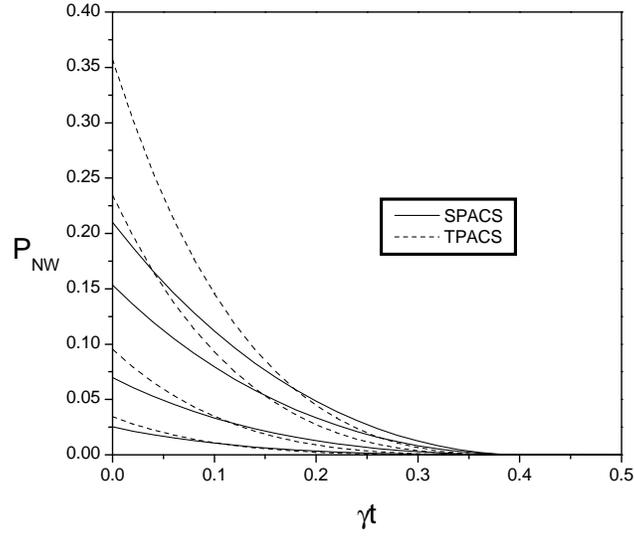}}
\caption{$P_{NW}$ is plotted as the function of the decay time
$\gamma{t}$ for SPACS and TPACS with different values of $\alpha$.
(Solid line) from top to bottom, SPACS with
$\alpha=0.1,0.5,1.0,1.5$, respectively; (Dash line) from top to
bottom, TPACS with $\alpha=0.1,0.5,1.0,1.5$, respectively.
$n=0.5$.}
\end{figure}
\newpage
\begin{figure}
\centerline{\includegraphics[width=8.3cm]{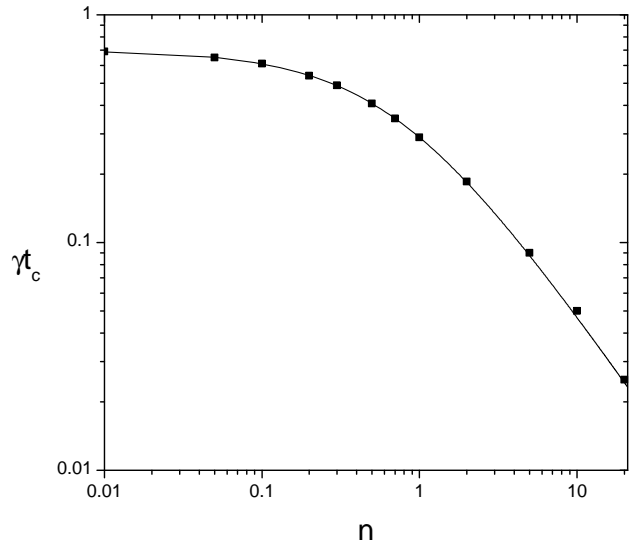}}
\caption{The threshold decay time $\gamma{t}_c$ beyond which
$P_{NW}=0$ is plotted as the function of the mean thermal photon
number $n$ for the case of the SPACS. (Solid square) numerical
results; (Solid line) $\gamma{t}_c=\ln\frac{2+2n}{1+2n}$. It is
shown that $\gamma{t}_c\simeq0.5/n$ when $n\gg1$. The numerical
calculations are based on the SPACS with $\alpha=0.5$.}
\end{figure}

\end{document}